\documentclass[twocolumn,prd,superscriptaddress,showpacs,floatfix,%
preprintnumbers, nofootinbib,hyperref]{revtex4}  
\usepackage{epsfig}   
\usepackage{bm}  
\usepackage{color}

\newcommand{\ben}{\begin{equation}}
\newcommand{\een}{\end{equation}}

\newcommand{\gtrsim}{\,\rlap{\lower3.7pt\hbox{$\mathchar\sim$}}
\raise1pt\hbox{$>$}\,}
\newcommand{\lesssim}{\,\rlap{\lower3.7pt\hbox{$\mathchar\sim$}}
\raise1pt\hbox{$<$}\,}

\newcommand{\be}{\begin{equation}}
\newcommand{\ee}{\end{equation}}  
\newcommand{\bea}{\begin{eqnarray}}
\newcommand{\eea}{\end{eqnarray}}

\def\theta{\vartheta}   

\begin{document}
\preprint{LAPTH-036/11}

\title{Instability in the dense supernova neutrino gas with \\  flavor-dependent angular distributions}

\author{Alessandro Mirizzi} 
\affiliation{II Institut f\"ur Theoretische Physik, Universit\"at Hamburg,
Luruper Chaussee 149, 22761 Hamburg, Germany} 

\author{Pasquale Dario Serpico} 
\affiliation{LAPTh, Univ. de Savoie, CNRS, B.P.110, Annecy-le-Vieux F-74941, France}


\begin{abstract}
The usual description of self-induced flavor conversions for neutrinos ($\nu$'s) in supernovae  is based
on the simplified assumption that all the $\nu$'s of the different species are emitted ``half-isotropically'' by
a common neutrinosphere, in analogy to a blackbody emission. However, realistic supernova simulations show
that $\nu$  angular distributions at decoupling are far from being half-isotropic and, above all, are flavor-dependent. We show that flavor-dependent angular distributions  may lead to crossing points in the angular spectra of different $\nu$ species (where $F_{\nu_e}=F_{\nu_x}$ and 
$F_{{\bar\nu}_e}=F_{{\bar\nu}_x}$) around which  a new multi-angle instability can develop. 
To characterize   this  effect, we carry out   a linearized  flavor stability analysis   for different
SN neutrino  angular distributions.
We find that this instability can  shift the onset of the flavor conversions toward  low-radii and  produce
a smearing of the splitting features found with trivial $\nu$ emission models. As a result the  spectral differences among $\nu$'s of different flavors
 could be strongly reduced.
\end{abstract}

\pacs{14.60.Pq, 97.60.Bw}   

\maketitle

\emph{Introduction.---}
The neutrino flux emitted  from a core-collapse supernova (SN)
represents a powerful tool to get valuable information about the  mixing parameters
and the dynamics of the exploding stellar core.
In this context, renewed attention is being paid
to collective features of flavor transformations~\cite{Duan:2006an,Hannestad:2006nj} induced by $\nu$-$\nu$ self-interactions~\cite{Pantaleone:1992eq,Qian:1994wh}
 in the deepest SN regions, near the neutrino-sphere (see~\cite{Duan:2010bg} for a recent review).
 In particular,  it has been claimed that the self-induced ``spectral splits'' 
would be observable in the  $\nu$ burst from the next galactic SN, 
allowing to get crucial information about the unknown $\nu$ mass ordering 
(see, e.g.,~\cite{Duan:2007bt}).

The development of these self-induced effects crucially depends  
 on the inner  boundary conditions fixed
for the flavor evolution, e.g. on the $\nu$ initial energy and angular distributions.
In this context,    
 all the recent numerical simulations  have 
been based on the so-called ``bulb model'' (see, e.g,~\cite{Duan:2006an,Fogli:2007bk}).
Assuming that the $\nu$ trapping region at high matter density can be decoupled from the region 
at lower density where flavor conversions
would start,  $\nu$'s of different species are considered as 
emitted ``half-isotropically'' (i.e. with  all outward-moving angular modes 
equally occupied and all the backward-moving modes
empty) by a common spherical ``neutrinosphere,''
in analogy with a blackbody emission. 
 However, this simplified toy-model may not  capture important features of 
the SN
$\nu$ emission.
In particular,  the transition between the isotropic $\nu$ emission in   trapping regime (at a higher matter density) to
the forward-peaked free streaming (at a lower matter density) does not necessarily imply a radius where the $\nu$ angular
distributions are half-isotropic (see, e.g.,~\cite{Ott:2008jb,Sarikas:2011am}). Moreover, the different $\nu$ species decouple
at different radii.
Therefore, the physical last-scattering neutrinospheres for the different species would not coincide.
 Then,  fixing a    common neutrinosphere as inner boundary for the flavor evolution,   one would
realistically find there different angular distributions
  for  the different $\nu$ species. In particular, since the ${\overline\nu}_e$'s and $\nu_x$'s  decouple at smaller radii with respect to $\nu_e$'s their 
distributions are more forward-peaked. 

The presence of non-trivial   angular distributions was claimed in~\cite{Sawyer:2005jk}   to produce 
a  multi-angle instability in the self-induced flavor evolution of a toy model of $\nu$ gas. 
However, in that seminal paper the author warned readers not to draw firm conclusions for  the realistic SN $\nu$ case
from his analysis performed with a small number of angular modes, before having explored this case with large-scale
numerical simulations. Stimulated also by his intriguing hint, we decided to begin a systematic study of 
this issue.
We find that as long as flavor
 universality is not broken in the $\nu$ angular distributions, only minor quantitative effects appear  in the evolution with respect to the
half-isotropic case.
Conversely,  flavor-dependent angular distributions  can dramatically affect the conversions of the dense SN $\nu$ gas. 
 We relate this behavior 
  to the presence of  crossing points in the  angular spectra of different $\nu$  species (i.e., at emission angles for which $F_{\nu_e}=F_{\nu_x}$ and 
$F_{{\bar\nu}_e}=F_{{\bar\nu}_x}$).
It is known that $\nu$ energy spectra presenting crossing points in the energy variable can develop instabilities around them~\cite{Dasgupta:2009mg}. This effect has been analytically understood performing a linearized stability analysis of the 
$\nu$ equations of motion~\cite{Banerjee:2011fj}.
Here, applying the same analysis,  we find that an analogous effect occurs when the angular spectra of different flavors present  crossing points.
As a consequence, this  multi-angle instability can lead to 
a wash-out of the splitting features found assuming  a half-isotropic $\nu$ emission.

\begin{figure}[!t]
\begin{center}  
\includegraphics[width=1.\columnwidth]{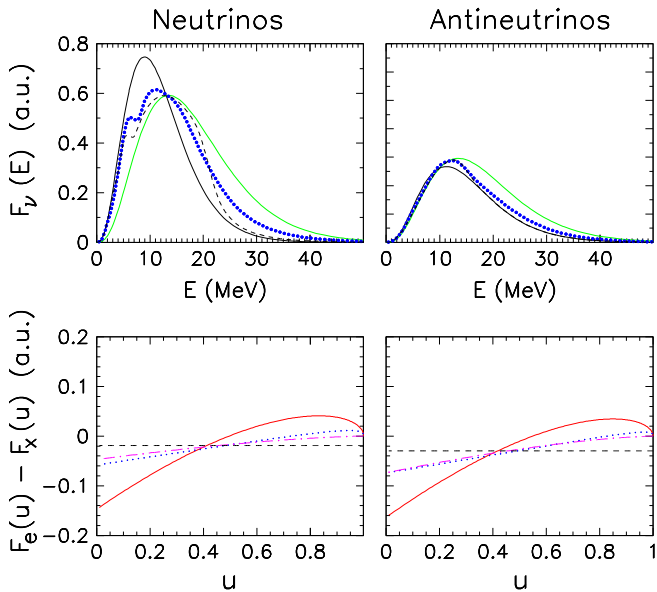}  
\end{center} 
\vspace{-0.4cm} 
\caption{Upper panels: Multi-angle flavor evolution  for $\nu$'s (left panel) and $\overline\nu$'s (right panel). 
Energy spectra initially for $\nu_e$ (black continuous curves) and $\nu_x$ (light continuous curves) and after collective oscillations
for $\nu_e$ with half-isotropic angular distribution (dashed curves) and for an angular distribution
with  $\beta_e=1.0$, $\beta_x=1.5$ (thick dotted curves).  
Lower panels: Difference of the energy-integrated angular spectra $F_{\nu_e}(u) -
F_{\nu_x}(u)$ for $\nu$'s (left panel) and $\overline\nu$'s (right panel) at the neutrinosphere
for the half-isotropic case $\beta_e=\beta_x=0.0$ (dashed curves) and for  the cases  with $\beta_e=1.0$, $\beta_x=1.5$ (dotted curves), $\beta_e=1.0$, $\beta_x=3.0$  (continuous curves) and   $\beta_e=\beta_x=3.0$  (dash-dotted curves).
\label{fig1}}  
\end{figure}  

\emph{Energy and angular distributions.---} 
In order to perform our study of the multi-angle instability we have to fix our $\nu$ emission models.
Our main goal would be to remove the bulb-model approximation of a half-isotropic neutrino angular distribution at the neutrinosphere,
equal for all the species. 
In order to introduce  $\nu$ angular distributions, it is  convenient to parameterize every angular mode in terms of its emission angle $\theta_R$ relative to the radial direction
of the neutrinosphere, that here we schematically fix at $R=10$~km. 
A further simplification is obtained if one labels the different angular modes in terms of the variable $u=\sin^2\theta_R$,
as in~\cite{EstebanPretel:2007ec,Banerjee:2011fj}. Note that 
for an half-isotropic emission at the neutrinosphere  the $\nu$ angular distribution of the radial fluxes
 is a box spectrum in $0\leq u \leq 1$~\cite{EstebanPretel:2007ec}.
 
We remark that energy and angular distributions of SN $\nu$'s  entering the $\nu$ number fluxes $F_{\nu_\alpha}(E,u)$, integrated over a sphere of radius $r$, are not independent
of each other. However, here we schematically  assume that the angular distributions are energy
independent. Then, we can factorize the $\nu$ flux of each flavor as  
$F_{\nu_\alpha}(E,u)= N_{\nu_\alpha} \times \varphi_{\nu_\alpha}(E) \times
U_{\nu_\alpha}(u).$
The function $\varphi_{\nu_\alpha} (E)$ is the normalized $\nu$ energy spectrum ($\int dE \varphi_{\nu_\alpha} (E)=1$),
which we will parametrize for the different flavors with  deformed Maxwell-Boltzmann distributions, as suggested by SN simulations~\cite{Raffelt:2003en}. 
The function $U_{\nu_\alpha}(u)$ is the normalized angular distribution ($\int du U_{\nu_\alpha}(u)$=1). 
We fix the neutrino average energies at
$ (\langle E_{\nu_e}\rangle, \langle E_{{\bar\nu}_e}\rangle, \langle E_{\nu_x}\rangle)=(12, 15, 18)\,\  
\textrm{MeV}$.
Concerning the possible $\nu$ flux ordering we consider a 
case representative of the cooling phase~\cite{Mirizzi:2010uz},
i.e. $N_{\nu_e}:N_{\bar\nu_e}:N_{\nu_x}=1.13:1.00:1.33$.
According to this choice, in the upper panels of Figure~1 are represented the initial 
 fluxes for $\nu_e$ (continuous black curves) and $\nu_x$ (continuous light curves) for
$\nu$'s (left panels) and $\bar\nu$'s (right panels) respectively.

A systematic characterization  of  the $\nu$ angular distributions at different radii in SNe is lacking in literature.
Therefore, inspired by an inspection
of some real angular distributions~\cite{supplement}, 
 we use a simple toy model to capture the main deviations with respect to the half-isotropic
bulb model, 
where $U_{\nu_\alpha}= 1$ for all the $\nu$ species. In particular,  we choose forward-peaked distributions
$U_{\nu_\alpha}(u) \propto (1-u)^{\beta_\alpha/2}$. For simplicity in the following  we  assume $U_{\nu_e}=U_{{\bar\nu}_e}$.
In the lower  panels of Figure~1 we plot   the energy-integrated spectral  difference $F_{\nu_e}(u)-F_{\nu_x}(u)$ for $\nu$'s (left panels) and the analogous one for $\overline\nu$'s (right panels)
for four representative $(\beta_e,\beta_x)$ cases.  Namely we compare  
the half-isotropic case  
($\beta_e=\beta_x=0$,  dashed curves)  with  three non-trivial angular distributions:
 $\beta_e=1.0$, $\beta_x=1.5$ (dotted curves), $\beta_e=1.0$, $\beta_x=3.0$ (continuous curves), and $\beta_e=\beta_x=3.0$ (dash-dotted curves).
One  realizes that
in the  half-isotropic case $\beta_e=\beta_x=0.0$ and in the flavor universal case
$\beta_e=\beta_x=3.0$, the differences of angular spectra do not present any crossing point in the angular
variable, while in  the other two cases, the spectra present a crossing point at finite $0<u<1$, where
 $F_{\nu_e}(u)=F_{\nu_x}(u)$. 
We will show  that crossing points
 in the energy-integrated angular spectra will produce a significant speed-up of the multi-angle instability  
 in the $\nu$ flavor conversions.

\emph{Setup of the flavor evolution.---}
Our description of the non-linear $\nu$ flavor conversions is based on a
two-flavor oscillation scenario. For simplicity, we refer to the cases
studied in~\cite{Mirizzi:2010uz} in  which three-flavor effects associated with
the solar sector are sub-leading. In this situation, self-induced oscillations 
are driven by the atmospheric mass-square difference
$\Delta m^2_{\rm atm} = m_3^2 - m_{1,2}^2 \simeq 2.0 \times 10^{-3}$~eV$^{2}$
and by a small (matter suppressed) in-medium mixing
$\Theta_{\rm m} = 10^{-3}$~\cite{Kuo:1989qe}.
We  refer to late-time cooling phase ($t\gtrsim 1$~s) where the electron density 
$n_e$ is smaller than the $\nu$ one $n_\nu$ and then has a subleading role on the development
of the collective oscillations~\cite{Chakraborty:2011gd}.   
Apart from the mixing suppression, we neglect the sub-leading matter effects in the flavor evolution. 
We will refer to the inverted mass hierarchy ($\Delta m^2_{\rm atm} <0$).

Following~\cite{Banerjee:2011fj}, we write the equations of motion for the flux matrices  $\Phi_{E,u}$  as function of the radial coordinate.
The diagonal $\Phi_{E,u}$ elements are
the ordinary number fluxes $F_{\nu_{\alpha}}(E,u)$.
We normalize the flux matrices to the total ${\overline\nu}_e$ number flux $N_{{\bar\nu}_e}$ at the neutrinosphere.
 Conventionally, we use negative $E$ and negative
number 
fluxes for $\bar\nu$'s. The off-diagonal elements,
which are initially zero, carry a phase information due to 
flavor mixing.
Then, the equations of motion read
$
\textrm{i}\partial_r \Phi_{E,u}=[H_{E,u},\Phi_{E,u}] \,\ 
\label{eq:eom1} $
with the  $\nu$-$\nu$ Hamiltonian~\cite{Pantaleone:1992eq,Qian:1994wh,Banerjee:2011fj}
\begin{equation}
H_{\nu\nu} = \frac{\sqrt{2}G_F}{4\pi r^2}\int_{-\infty}^{+\infty}d E^\prime
 \int_{0}^{1}du^\prime \left(\frac{1-v_{u}v_{u^\prime}}{v_{u}v_{u^\prime}}
 \right)\Phi_{E^\prime,u^\prime} \,\, .
 \label{eq:eom2}
\end{equation}

The 
 factor proportional to the neutrino velocity $v_{u,r} = (1-u R^2/r^2)^{1/2}$~\cite{EstebanPretel:2007ec} 
implies ``multi-angle'' effects for neutrinos moving on different trajectories~\cite{Pantaleone:1992eq,Qian:1994wh, Duan:2006an}. In order to properly simulate numerically this effect one needs 
to  follow a large number
$[{\mathcal O}(10^3)]$ of interacting $\nu$ modes. 

\emph{Stability condition.---}
In order to achieve a deeper understanding of the multi-angle instability, triggered by non-trivial
angular distributions,  we find particularly useful to apply to our problem the linearized stability analysis
recently worked out in~\cite{Banerjee:2011fj}. This method  allows us to determine the onset of the flavor conversions, seeking
an exponentially growing solution in the eigenvalue equations, associated with the linearized equations of motion for
the $\nu$ ensemble.

In order to perform the stability analysis we closely follow the
prescriptions presented in~\cite{Banerjee:2011fj} and summarized in the following. 
At first we switch to the 
frequency variable $\omega= \Delta m^2_{\rm atm}/2E$ and  we introduce the 
neutrino flux difference distributions $g_{\omega,u}\equiv g(\omega,u) =
|d\omega/dE|[F_{\nu_e}(E(\omega),u)-F_{\nu_x}(E(\omega),u)]$, 
normalized  to the total ${\overline\nu}_e$ flux at the neutrinosphere. 
Note that  $g_{\omega,u}$ is defined also for negative $\omega$, where it represents the difference of fluxes in the antineutrino sector in the opposite ordering. 
Then, we   write the flux matrices as~\cite{Banerjee:2011fj}
\begin{equation}
\Phi_{\omega,u}= \frac{\textrm{Tr}\Phi_{\omega,u}}{2}+
\frac{g_{\omega,u}}{2}
\left( \begin{array}{cc} s_{\omega,u} &  S_{\omega,u} \\
S^{\ast}_{\omega,u} & -s_{\omega,u} 
\end{array} \right) \,\ ,
\end{equation}
where $\textrm{Tr}\,\Phi_{\omega,u}$ is conserved and then irrelevant for the flavor conversions, and the initial conditions for the  
``swapping matrix'' in the second term on the right-hand side are $s_{\omega,u}=1$ and $S_{\omega,u}=0$.
Self-induced flavor transitions  start when the off-diagonal term $S _{\omega,u}$
 exponentially grows.
In the small-amplitude limit $|S _{\omega,u}|\ll 1$, 
and at far distances from the neutrinosphere $r \gg R$,
the linearized evolution equations for  $S _{\omega,u}$
lead to an eigenvalue equation~\cite{Banerjee:2011fj}
 \begin{equation}
 (\omega+u\epsilon\mu-\Omega) Q _{\omega,u} = \mu \int du^\prime d\omega^\prime (u+u^\prime)
g_{\omega^\prime, u^\prime} Q _{\omega^\prime,u^\prime} \,\ ,
\label{eq:eigen}
 \end{equation}
for the eigenvector $Q _{\omega,u}$, obtained writing the solution of the linearized equation $S _{\omega,u} = Q _{\omega,u} e^{-i\Omega r}$.
Here we introduced the complex frequency
$\Omega= \gamma + i \kappa$ and
the parameter $ \epsilon = \int du \,\ d\omega \,\ g_{\omega,u}$
quantifying the ``asymmetry'' of the neutrino spectrum, normalized to the total ${\overline\nu}_e$ 
number flux. For our specific choice 
of $\nu$ spectra, it results $\epsilon=0.13$.
The $\nu$-$\nu$ interaction strength is given by
\begin{equation}
\mu = \frac{\sqrt{2}G_F N_{{\bar\nu}_e}}{4 \pi r^2}\frac{R^2}{2 r^2} \,\ .
\end{equation}
 For our SN model, it results $\mu(R) = 4.7 \times 10^{-4}$~km$^{-1}$.
A solution of Eq.~(\ref{eq:eigen}) with 
$\kappa >0$ would indicate an exponential increasing $Q _{\omega,u}$, i.e. an instability.
When an instability occurs, for a given  angular mode $u_0$ the function 
$|Q _{\omega,u_0}|$ is a Lorentzian~\cite{Dasgupta:2009mg}, centered around a \emph{resonance frequency}
$\omega = \gamma - u\epsilon \mu,$
and with a width $\kappa$.

\emph{Flavor conversions---}
 In order to illustrate the effect of the angular distributions,  in Fig.~1 (upper panels) we compare the oscillated
electron (anti)neutrino fluxes (at $r=300$~km) in the half-isotropic
case ($\beta_e=\beta_x=0$; dashed curve) with a non-trivial angular distribution
 with $\beta_e=1.0$, $\beta_x=1.5$ (thick dotted curve).  
The effect on the final spectra is dramatic: 
The spectral splits observable in the final ${\nu}_e$ spectra in the half-isotropic case are smeared-out
and the spectral swaps  are not complete.

\begin{figure}  
\includegraphics[angle=0,width=1.\columnwidth]{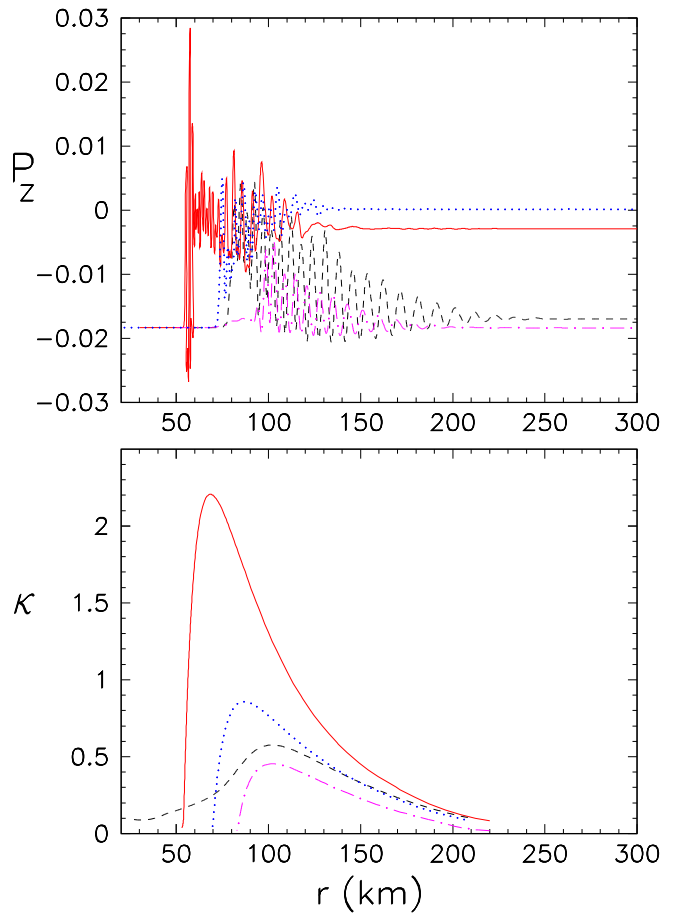}
\vspace{-0.4cm}
\caption{Upper panel: Radial evolution of the integrated z-component
of the polarization vector $P_z$ for $\nu$'s for the half-isotropic case  
(dashed curve) and for angular distributions
with $\beta_e=1.0$, $\beta_x=1.5$ (dotted curve), $\beta_e=1.0$, $\beta_x=3.0$ (continuous curve), and 
$\beta_e=\beta_x=3.0$ (dot-dashed curve). Lower-panel: Radial evolution of the eigenvalue $\kappa$ for the same cases shown in
the upper panel. 
\label{fig2}} 
\end{figure}  

In the upper panel of Fig.~\ref{fig2} we show
the radial evolution of the z-component of
the integrated value
of the $\nu$ polarization vector $P_z$,
that is related to the flavor content of the ensemble
for four different angular distributions of Fig.~\ref{fig1}. Namely,
we compare  the half-isotropic case  
($\beta_e=0$, $\beta_x=0$,  dashed curve) with  three non-trivial angular distributions:
 $\beta_e=1.0$, $\beta_x=1.5$ (dotted curve), $\beta_e=1.0$, $\beta_x=3.0$ (continuous curve), and 
$\beta_e=\beta_x=3.0$ (dot-dashed curve).
 First of all, note that with the  $\beta_e=\beta_x=3.0$ flavor-blind distributions the behavior does not present major changes with respect to the naive case, except from a delay in the starting of  flavor conversions at
$r\simeq 90$~km. 
 In the case of $\beta_e=1.0$, $\beta_x=1.5$,  the onset of the flavor conversions
(at $r\simeq 70$ km)
is very close to what expected in the half-isotropic case ($r\simeq 75$ km).  
However, the further flavor evolution shows dramatic differences. Indeed, with the non-trivial
angular distribution the final value of the integrated $\nu$ polarization vector would be $P_z \simeq 0$ indicating
that the spectral differences between the final $\nu_e$ and $\nu_x$ spectra are strongly
reduced with respect to the half-isotropic case. 
Choosing larger differences in the angular distributions of different flavors, as in the case with
$\beta_e=1.0$, $\beta_x=3.0$, not only the final value of the $P_z$ will change, but also  
flavor conversions would start much earlier ($r \simeq 55$ km) than in the half-isotropic case. 

In the bottom panel of 
Fig.~\ref{fig2} we plot the radial evolution of the eigenvalue $\kappa$ coming 
from the solution of Eq.~(\ref{eq:eigen}) for the same cases shown in the upper panel. 
In the  half-isotropic case $\beta_e=\beta_x=0$ (dashed curve) the $\kappa$ function presents
a hump peaked around $r \simeq 100$~km and connected with a long tail around $r \simeq 75$~km. 
This tail indicates that the system is in principle  always unstable. However, one can verify 
that in the tail the unstable $\nu$ frequency modes 
are in the infrared region, 
where the  spectrum is strongly suppressed. Therefore, they have no 
impact for the flavor conversions. 
The onset of the flavor conversions is then given by the connection between the hump and the tail
 at $r=75$~km. This result is  in agreement with what shown in the numerical calculation
of $P_z$, shown in the upper panel (dashed curve).

When considering non-trivial angular distributions, the long tail in the
$\kappa$ function in the half-isotropic case now disappears. 
However, the presence of a non-isotropic angular distribution is not enough to produce an enhancement in the value of
 $\kappa$. 
Conversely, in the flavor blind case $\beta_e=\beta_x=3.0$ (dot-dashed curve) the instability is suppressed with respect to what 
has been seen in the half-isotropic case. 
 In this case, flavor conversions start at $r=80$~km. 
  A significant enhancement of the multi-angle instability occurs when $\nu$ angular spectra exhibit crossing points 
 in $u$, as can be seen in the case  with $\beta_e=1.0$, $\beta_x=1.5$ (dotted curve)  and even more in the one
 with $\beta_e=1.0$,  $\beta_x=3.0$ (continuous curve).
 In both  cases the peak in $\kappa$ increases with respect to the half-isotropic case and 
 also the hump broadens toward smaller $r$.
Since $\kappa$ reaches a higher peak value, 
the width of the Lorentzian around an unstable frequency mode for a given angle $u$ would be broad,
implying a speed-up in the transitions.  
 In particular, in the case of $\beta_e=1.0, \beta_x=1.5$ flavor conversions  start
at $r\simeq 70$~km, while in the case with  $\beta_e=1.0, \beta_x=3.0$
 around 
$r \simeq 55$~km, in agreement
with the numerical results shown in the upper panel of Fig.~\ref{fig2}.

\emph{Conclusions---}
We have described a new instability in self-induced flavor conversions for SN $\nu$'s, associated with flavor-dependent 
angular distributions. These can lead to crossing points among the different spectra that can produce new 
flavor conversion effects, absent with an half-isotropic   $\nu$ emission. 
We checked that this effect would develop in both the mass hierarchies and would be particularly relevant during the SN
cooling phase, where the differences among the different flavors are relatively small, and self-induced effect can 
develop without any matter hindrance. 
The effect on the oscillated $\nu$ spectra can be dramatic. Namely, it would  produce a smearing of the splitting features widely discussed in the half-isotropic case, and resulting $\nu$ fluxes with less significant spectral differences. This tendency toward spectral equalization would
challenge the detection of further oscillation signatures, like the ones associated with the
Earth crossing of SN $\nu$'s (see, e.g.,~\cite{Lunardini:2001pb}). 
Moreover, we found that also onset of the flavor conversions can be significantly pushed at low-radii, challenging the
multi-angle suppression found in the half-isotropic case~\cite{Duan:2010bf}. 
Possible flavor conversions at low-radii would have an interesting impact on the r-process
nucleosynthesis in SNe~\cite{Duan:2010af}. Furthermore, if $\nu$ oscillations develop too close to the neutrinosphere, they would invalidate the $\nu$ transport paradigm in SNe that ignores  $\nu$ conversions. 
Given the importance of the effects discussed here, a  future task would be to
perform the  flavor evolution and stability analysis  taking into account angular distributions as realistic
as possible, obtained directly from SN  simulations. 

\vspace{-0.6cm}

\section*{Acknowledgements} 
\vspace{-0.5cm} 
We thank  G.~ Raffelt for useful comments.
The work of A.M.  was supported by the German Science Foundation (DFG)
within the Collaborative Research Center 676 ``Particles, Strings and the
Early Universe''.


\end{document}